\documentclass{article}
\usepackage{spconfa4,amsmath,graphicx}
\usepackage{units}
\usepackage[colorlinks=false,pdfborder={0 0 0}]{hyperref}

\title{Long-term Conversation Analysis: Privacy-Utility Trade-off under Noise and Reverberation}
\name{Jule Pohlhausen$^{1,2}$, Francesco Nespoli$^{3}$, Joerg Bitzer$^{1,4}$\thanks{The authors would like to thank Volker Hohmann for his helpful comments on the article.
This work was supported by the Graduation program of Jade University of Applied Sciences (Jade2Pro 2.0) and by the European Union Horizon 2020 program under the Marie Sk\l{}odowska-Curie grant No 95636.}}
\address{
$^1$Jade University of Applied Sciences, Institute of Hearing Technology and Audiology, Oldenburg, Germany\\
$^2$Carl von Ossietzky Universität Oldenburg, Dept. Medical Physics and Acoustics, Germany\\
$^3$Microsoft, London, UK
and Imperial College, Dept. Electrical and Electronic Engineering, London, UK\\
$^4$Fraunhofer IDMT Dept. HSA, Oldenburg, Germany
}
\begin{document}
%\ninept
%
\maketitle
\begin{abstract}
%100 to 150 words
Recordings in everyday life require privacy preservation of the speech content and speaker identity. This contribution explores the influence of noise and reverberation on the trade-off between privacy and utility for low-cost privacy-preserving methods feasible for edge computing. These methods compromise spectral and temporal smoothing, speaker anonymization using the McAdams coefficient, sampling with a very low sampling rate, and combinations.
Privacy is assessed by automatic speech and speaker recognition, while our utility considers voice activity detection and speaker diarization. 
Overall, our evaluation shows that additional noise degrades the performance of all models more than reverberation. This degradation corresponds to enhanced speech privacy, while utility is less deteriorated for some methods.
\end{abstract}
\begin{keywords}
privacy, conversation analysis, speech recognition, speaker recognition, voice activity detection, speaker diarization
\end{keywords}

\section{Introduction} \label{sec:intro}
The conversational behavior in our everyday life is valuable in predicting sociability and health outcomes \cite{ milek_eavesdropping_2018}. %harari_sensing_2020,
This contribution focuses on tools for analyzing long-term recordings in everyday life, captured by portable devices over several days. 
In particular, the objective analysis of conversations requires a robust detection of voiced time segments (voice activity detection, VAD), which are subsequently attributed to different speakers (speaker diarization, SD). These steps reveal speaker turns and the involvement of different communication partners \cite{wyatt2007conversation}.
Since speech contains important personal identifiable information (PII) and real-life recordings involve uncontrolled environments, both in private and public, recordings in everyday life require privacy preservation and the protection of speech data, as stated in the European general data protection regulation (EU GDPR). 
Hence, our goal is to record only acoustic features capable of analyzing conversations while protecting the two main privacy aspects of speech data: the linguistic speech content and the speaker identity \cite{nelus2016towards}. These requirements define the trade-off between privacy and utility.

Furthermore, the computational power, memory, and power consumption of portable recording devices demand features with low computational complexity.
The described limitations imposed by portable devices conflict with the use of existing privacy-preserving methods like encrypting speech signals \cite{seven} or performing federated learning \cite{federated}. 
The conversation detection and SD system proposed by \cite{wyatt2007conversation} is based on entropy, correlation, and energy features to preserve speech privacy. However, it requires one audio stream per speaker and a central node to combine the information from all streams, which is not feasible for unconstrained recording environments. 
Alternatively, the Electronically Activated Recorder (EAR) records intermittently snippets of audio, e.g., 30~s or 50~s recordings every 9~min or 12.5~min \cite{mehl2017ear}. Since uninvolved parties could be recorded without their consent, this approach can not be applied in public environments.
In contrast, this contribution focuses on low-cost methods such as spectral and temporal smoothing \cite{bitzer2016privacy}, speaker anonymization using the McAdams coefficient \cite{patino2021mcadams}, and low-frequency audio \cite{raman2022conflab}.

The privacy evaluation considers the word error rate (WER) of an automatic speech recognition (ASR) model and the equal error rate (EER) of an automatic speaker verification (ASV) system. The utility performance is assessed with Matthews correlation coefficient (MCC) of a VAD model and the diarization error rate (DER) extracted from a SD system, which assigns unique speaker labels that are not linked to the speaker's identity.
We aim to maintain utility performance while enforcing privacy, therefore minimizing DER while maximizing MCC, WER, and EER. Presently, most assessments rely on optimal testing conditions \cite{voicepriv2020}. Hence, this contribution investigates the impact of noise and reverberation on the trade-off between privacy and utility.

%The remainder of this contribution is organized as follows: First, Section~\ref{sec:pripre} introduces our considered privacy-preserving features. The experimental setup is described in Section~\ref{sec:setup} our evaluation results are presented and discussed in Section~\ref{sec:results}. Finally, Section~\ref{sec:con} offers conclusions and future research directions.

%%%%%%%%%%%%%%%%%%%%%%%%%%%%%%%%%%%%%%%%%%%%%%%%%%%%%%%%%%%%%%%%%%
\section{Privacy-preserving methods} 
\label{sec:pripre}%
%%%%%%%%%%%%%%%%%%%%%%%%%%%%%%%%%%%%%%%%%%%%%%%%%%%%%%%%%%%%%%%%%%
Our main assumption is that all models rely on the same input features, namely log Mel filterbank energies computed on time segments of \unit[25]{ms} with a hop size of \unit[$L = 10$]{ms}.
This section summarizes the considered privacy-preserving methods applied during feature extraction or as pre-processing in the case of the McAdams anonymization \cite{patino2021mcadams}.%\\

\subsection{Spectral smoothing}\label{sec:smooth}
%\noindent \textbf{Spectral smoothing}\\
% Spectral Smoothing: Mel filter reduction
Analogously to reducing visual information content by pixelating images, mosaic speech corresponds to pixelated spectrograms that were smeared in the time-frequency domain and decrease the speech intelligibility \cite{nakajima2018mosaic_speech}. 
Hence, in our case, spectral smoothing corresponds to decreasing the number of filters of the Mel filterbank while covering the same frequency range. We evaluated filterbanks with 80 Mel filters, which refers to the baseline performance, and 10 Mel filters. %\\

\subsection{Temporal smoothing}
%\noindent \textbf{Temporal smoothing}\\
Similarly to mosaic speech \cite{nakajima2018mosaic_speech}, smoothed and subsampled power spectral densities (PSD) lead to degraded speech intelligibility \cite{bitzer2016privacy}.
These PSD were calculated with a window size of \unit[25]{ms} and a hop size of \unit[$L = 12.5$]{ms}. The PSD were smoothed with a first-order recursive filter with a smoothing time $\tau$ and subsampled by a factor of $\tau/L$, which corresponds for \unit[$\tau = 125$]{ms} to a subsampling factor $\tau/L = 10$, i.e., only every tenth smoothed PSD is stored. To restore the original time resolution of the frames, the smoothed and subsampled PSD were repeated by the subsampling factor before applying the Mel filterbank. The evaluation considers \unit[$\tau = 125$]{ms}, \unit[250]{ms}, \unit[375]{ms} with corresponding subsampling factors.%\\

\subsection{McAdams speaker anonymization}\label{sec:speaker_anon}
%\noindent \textbf{McAdams speaker anonymization}\\
%Our speaker anonymization is based on low-cost signal processing techniques feasible for resource-limited recording devices. 
The McAdams coefficient shifts the formant positions to hide speaker-related information while preserving the spoken content \cite{patino2021mcadams} and was the secondary baseline of the Voice Privacy Challenge (VPC) \cite{voicepriv2020}.
For semi-informed attackers, who have complete knowledge about the anonymization system, sampling a McAdams coefficient between 0.5 and 0.9 leads utterance-wise to strong privacy preservation in terms of EER \cite{patino2021mcadams}, but speaker-wise only to modestly deteriorated EER \cite{voicepriv2020} compared to unprotected signals.
Since the McAdams coefficient was designed to preserve ASR, it deteriorated the WER only marginally \cite{voicepriv2020, patino2021mcadams}. 
Therefore, to further investigate privacy preservation, 
we combined the McAdams coefficient with spectral and temporal smoothing.%\\

\subsection{Low-frequency audio}
%\noindent \textbf{Low-frequency audio}\\
The wearable badges in \cite{raman2022conflab} record low-frequency audio with a sampling rate of $f_s=1250$~Hz %and are combined with aerial-view video recordings 
to collect privacy-preserving %multimodal 
data of social conversations.
For evaluating this method, frequencies above $f_s/2=625$~Hz were discarded.

%%%%%%%%%%%%%%%%%%%%%%%%%%%%%%%%%%%%%%%%%%%%%%%%%%%%%%%%%%%%%%%%%%
\section{Experimental setup}
\label{sec:setup}%
%%%%%%%%%%%%%%%%%%%%%%%%%%%%%%%%%%%%%%%%%%%%%%%%%%%%%%%%%%%%%%%%%%
For assessing privacy and utility, this section describes the evaluation metrics, datasets, and augmentation.
This contribution considers \textit{semi-informed} attackers, i.e., the attacker has complete knowledge about the utilized privacy-preserving method and retrains or fine-tunes models on processed data.
Following the VPC \cite{voicepriv2020}, the semi-informed attacker is the strongest attack paradigm.
All experiments were conducted with SpeechBrain \cite{speechbrain} using two Nvidia GeForce RTX 3090 Ti GPU and one Intel Alder Lake CPU.
We report our results with confidence intervals, that were calculated with a confidence level $\alpha = 5$~\% using bootstrapping\footnote{https://github.com/luferrer/ConfidenceIntervals}. %\cite{confidence}.

\subsection{Privacy Assessment}
In this contribution, an ASR and an ASV model assess the privacy preservation of the speech content and the speaker identity, respectively.
%\subsubsection{Automatic Speech Recognition}\label{sec:ASR}
The ASR system is based on a transformer acoustic model encoder and a joint transformer decoder with connectionist temporal classification (CTC) \cite{CTC}, with the decoding stage integrating also CTC probabilities. The transcription performance is assessed with the WER, %defined as \begin{equation} \mathrm{WER} = \frac{N_{\mathrm{sub}}+N_{\mathrm{ins}}+N_{\mathrm{del}}}{N_{\mathrm{tok}}} \end{equation} where $N_{\mathrm{sub}},N_{\mathrm{ins}},N_{\mathrm{del}}$ are the number of substitutions, insertions, and deletions in the ASR output respectively, and $N_{\mathrm{tok}}$ is the number of tokens in the ground-truth transcript. Hence, 
where a high WER is desirable in order to protect the speech content.

%\subsubsection{Automatic Speaker Verification}\label{sec:ASV}
The ASV system uses the ECAPA-TDNN speaker encoder \cite{ECAPA} coupled with cosine similarity scoring for the verification task. The privacy evaluation relies on the EER, which is the operating point defined by the detection threshold at which the false acceptance rate and the false rejection rate are equal. Therefore, a higher EER indicates better anonymization and an EER of 50~\% reflects a random decision behavior. We report the weighted averaged EER on the VPC test subsets as in \cite{voicepriv2020}.

\begin{figure*}[ht!]
    \centering
    \includegraphics[width=2.07\columnwidth]{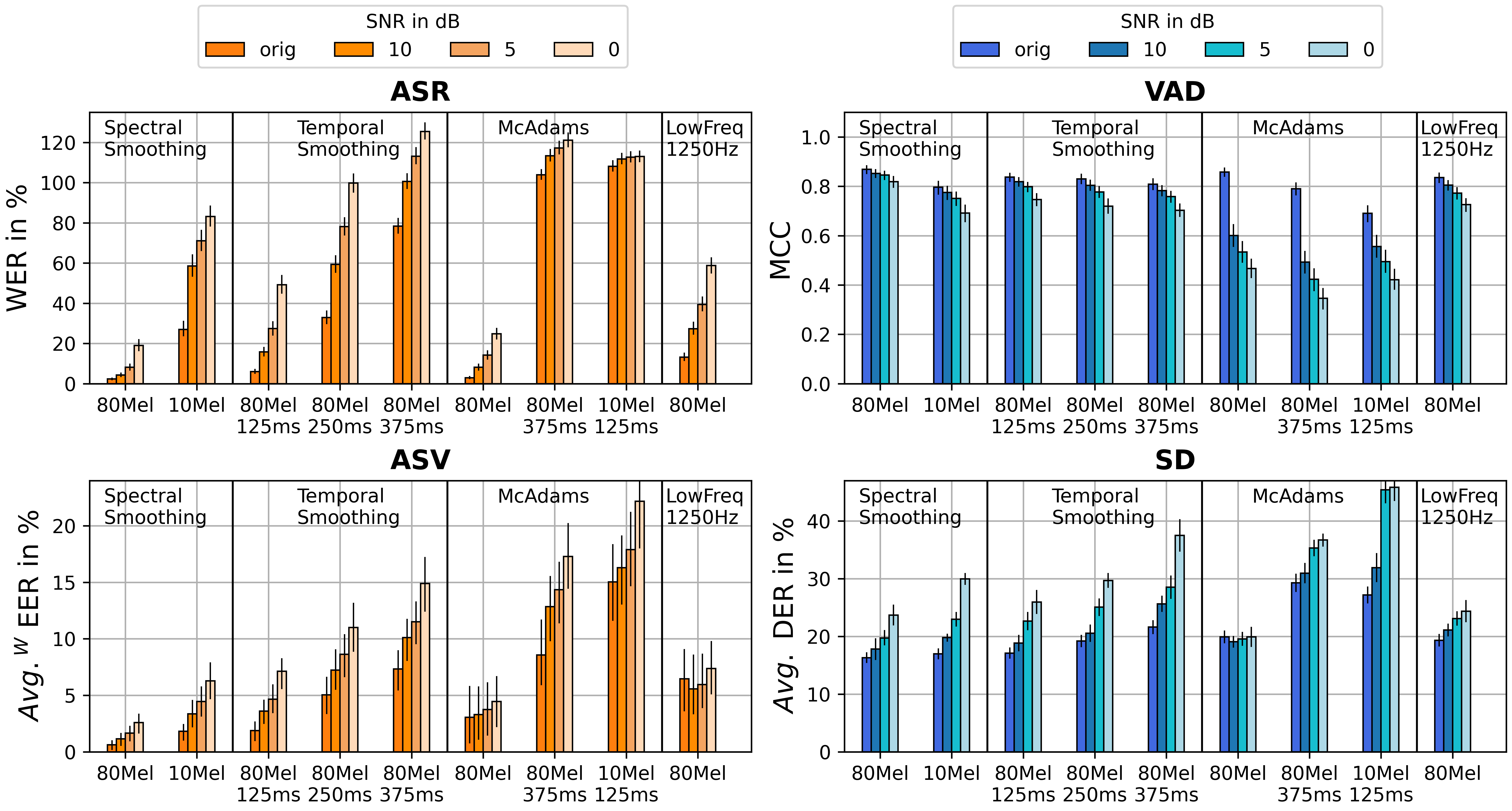}
    \caption{Privacy (ASR and ASV, orange) and utility (VAD and SD, blue) assessment for different SNR in dB, where the original test data is denoted by \textit{orig}. The evaluation considers spectral smoothing (Mel), temporal smoothing ($\tau$ in ms), McAdams anonymization \cite{patino2021mcadams}, and low-frequency audio with $f_s=1250$~Hz \cite{raman2022conflab}. The ASR is evaluated with the WER on the test-clean set of LibriSpeech \cite{librispeech}; the ASV is evaluated with the weighted averaged EER on the test sets of the VPC \cite{voicepriv2020}; the VAD is evaluated with the MCC on the test split of LibriParty \cite{speechbrain}; the SD estimates the number of speakers and is evaluated with the averaged DER on the eval set of the AMI Corpus \cite{carletta_unleashing_2007}. The confidence intervals are given by the vertical lines.}
    \label{fig:noise}
\end{figure*}

\subsection{Utility Assessment}
Our utility assessment considers a VAD and a SD model.
The VAD model consists of a convolutional recurrent deep neural network \cite{speechbrain} and performs binary classification, i.e., speech vs. non-speech, based on frame-level posterior probabilities which are processed with a sigmoid function. The VAD performance is evaluated with the MCC,
%\begin{equation} 
%\footnotesize
%\mathrm{MCC} = \frac{\mathrm{TP}\times \mathrm{TN} - \mathrm{FP}\times \mathrm{FN}}{\sqrt{(\mathrm{TP+FP})\cdot (\mathrm{TP+FN})\cdot (\mathrm{TN+FP})\cdot (\mathrm{TN+FN})}} 
%\end{equation}  
%\normalsize 
%here TP, TN, FP, and FN are the number of true positives, true negatives, false positives, and false negatives, respectively. The MCC  
which equals (-)1 for perfect (mis-)classification and 0 for random guessing.

%\subsubsection{Speaker Diarization}\label{sec:SD}
The SD model employs the same ECAPA-TDNN embeddings as the ASV model and performs Spectral Clustering (SC) to assign a relative speaker label to each time segment \cite{dawalatabad_ecapa-tdnn_2021}.  
For each meeting, the SD model estimates the number of speakers and utilizes oracle VAD information.
We report the averaged DER with a forgiveness collar of 250~ms.
%\begin{equation} 
%\mathrm{DER} = \frac{\mathrm{FA+MS+SER}}{\mathrm{TOTAL}} 
%\end{equation} 
%consisting of false alarms (FA), missed speech (MS), speaker confusion errors (SER), and the total reference speaker time (TOTAL) \cite{fiscus2006}.

\begin{figure*}[ht!]
    \centering
    \includegraphics[width=2.07\columnwidth]{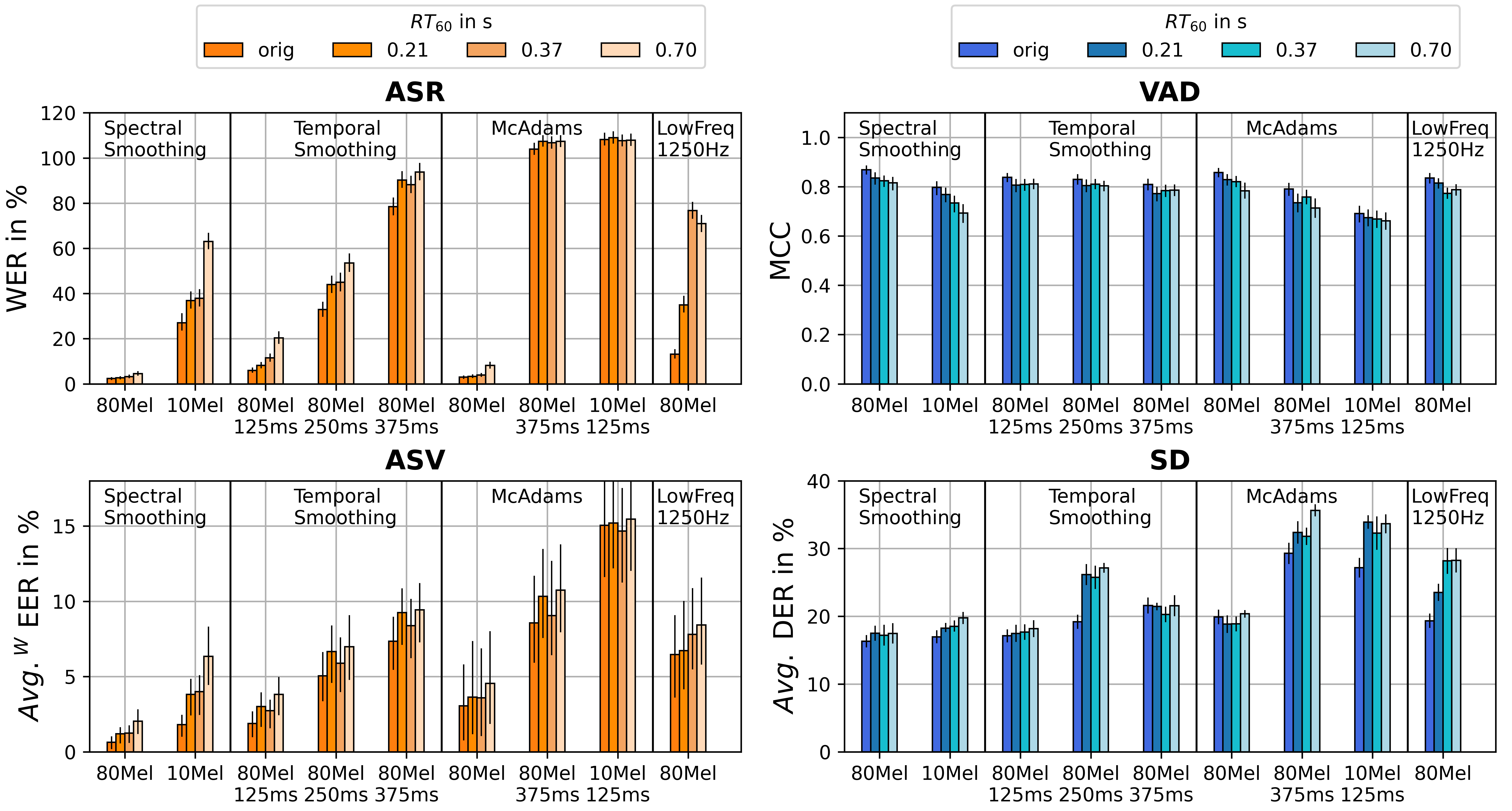}
    \caption{Privacy and utility assessment as in Figure~\ref{fig:noise}, but for different reverberation times $RT_{60}$ in s.}
    \label{fig:reverb}
\end{figure*}

\subsection{Databases}\label{sec:Data}
All experiments are performed on publicly available datasets sampled at 16~kHz. The ASR models were trained on the full 960 hours of LibriSpeech \cite{librispeech} and tested on the test-clean subset. For speaker verification and diarization, the ECAPA-TDNN speaker encoder \cite{ECAPA} was trained on VoxCeleb~2 \cite{VoxCeleb2}. 
The ASV performance was evaluated on the test splits of the VPC \cite{voicepriv2020}, while the SD performance was evaluated on the Augmented Multi-party Interaction (AMI) Meeting Corpus \cite{carletta_unleashing_2007} with the standard Full-corpus-ASR partition using the HeadsetMix recording streams. 
The VAD models were trained and tested on the train and test split of LibriParty \cite{speechbrain}.

For semi-informed attackers, an anonymized version of LibriSpeech 360~h \cite{librispeech} is obtained by randomly sampling a McAdams coefficient from $(0.5, 0.9)$ for each utterance.
This anonymized version was used to fine-tune the ASR model and the ECAPA-TDNN for further 30 epochs.
Moreover, the VAD model was fully retrained on the anonymized train split of LibriParty \cite{speechbrain}, since it was the least complex model. 
All fine-tuned and retrained models were tested on anonymized versions of the corresponding test data. 
During ASV evaluation, we randomly chose a McAdams coefficient per speaker for enrollment and trial data.

%\subsubsection{Noises}
For simulating typical environmental noise, 843 point-source noises that are sampled from the Freesound portion of the MUSAN corpus \cite{musan, ko2017augment} were added to the test audio signals at signal-to-noise ratios (SNR) of 10~dB, 5~dB, and 0~dB, which are typical values for realistic sound environments \cite{smeds_estimation_2015}.
%\subsubsection{Room Impulse Responses}

For simulating reverberation, we convolved the test data with three different room impulse responses (RIR), which were measured in a meeting, office, and lecture room and correspond to reverberation times $RT_{60}$ of 0.21~s, 0.37~s, and 0.70~s, respectively \cite{jeub2009binaural}.\\

\section{Results and discussion} \label{sec:results}
The results of our privacy and utility assessment with additional noise are presented in Figure~\ref{fig:noise} and with reverberation in Figure~\ref{fig:reverb}. Overall, for most methods in Section~\ref{sec:pripre} both, noise and reverberation, lead to a degraded performance for all models, where the influence of SNR tends to be higher compared to $RT_{60}$. For increasing SNR the evaluation metrics degrade, however, this deterioration is in some cases not significant as indicated by overlapping confidence intervals. 

For the ASR performance, the influence of noise and reverberation is most detrimental for methods that lead to a stronger information reduction, e.g., compare for spectral smoothing 80 and 10 Mel filters and for temporal smoothing $\tau$ of \unit[125]{ms} and  \unit[250]{ms}. Moreover, for methods with high WER on the original (orig.) test data, the additional noise or reverberation leads to ceiling effects, e.g., compare McAdams anonymization with 80 Mel filters and \unit[$\tau = 375$]{ms} or 10 Mel filters and \unit[$\tau = 125$]{ms}. Contrarily to all other methods, the ASR performance for low-frequency audio with $f_s=1250$~Hz is more degraded by increasing $RT_{60}$ compared to increasing the SNR.

The confidence intervals of the ASV performance seem to be higher in comparison to the remaining models, however, the different ranges and scales of the evaluation metrics must be considered. In general, the confidence intervals reflect the different levels of identity privacy for different speakers. The effect of reverberation is for most methods less significant than the intervariability between the speakers.

Regarding the utility, the VAD performance is relatively robust against noise and reverberation. However, for McAdams anonymization, the additional noise leads to the strongest deterioration compared to the original test data. Since McAdams anonymization is applied to the entire signal, thus also speech pauses are processed,  this seems to be more detrimental with additional noise.

Finally, the SD performance degrades with increasing SNR, except for the McAdams anonymization without additional smoothing. With reverberation, the SD performance appears to be robust for some methods compared to the original test data but degrades significantly for temporal smoothing with \unit[$\tau = 250$]{ms}, McAdams anonymization with additional smoothing, and low-frequency audio.

Overall, noise and reverberation enhance speech privacy as the ASR and ASV performance degrade. Simultaneously, the utility performance of VAD and SD also degrades, but depending on the privacy-preserving method this deterioration is less severe compared to the deterioration of the ASR and ASV performance.

\section{Conclusions} \label{sec:con}
This contribution investigated the influence of noise and reverberation on the trade-off between privacy and utility for privacy-preserving features feasible for edge computing.
Overall, the evaluation demonstrates that decreasing the SNR leads for most methods to stronger performance degradation compared to increasing $RT_{60}$. The ASR performance shows the strongest deterioration with increasing SNR, while the VAD is more robust, except for the combinations with McAdams anonymization.
Future work will focus on exploring the practical applicability and different input features beyond standard Mel filterbank energies to optimize both privacy and utility simultaneously.

% References should be produced using the bibtex program from suitable
% BiBTeX files (here: strings, refs, manuals). The IEEEbib.bst bibliography
% style file from IEEE produces unsorted bibliography list.
% -------------------------------------------------------------------------
\newpage
\bibliographystyle{IEEEtran}
\bibliography{literature}

% Generated by IEEEtran.bst, version: 1.13 (2008/09/30)
\begin{thebibliography}{10}
\providecommand{\url}[1]{#1}
\csname url@samestyle\endcsname
\providecommand{\newblock}{\relax}
\providecommand{\bibinfo}[2]{#2}
\providecommand{\BIBentrySTDinterwordspacing}{\spaceskip=0pt\relax}
\providecommand{\BIBentryALTinterwordstretchfactor}{4}
\providecommand{\BIBentryALTinterwordspacing}{\spaceskip=\fontdimen2\font plus
\BIBentryALTinterwordstretchfactor\fontdimen3\font minus
  \fontdimen4\font\relax}
\providecommand{\BIBforeignlanguage}[2]{{%
\expandafter\ifx\csname l@#1\endcsname\relax
\typeout{** WARNING: IEEEtran.bst: No hyphenation pattern has been}%
\typeout{** loaded for the language `#1'. Using the pattern for}%
\typeout{** the default language instead.}%
\else
\language=\csname l@#1\endcsname
\fi
#2}}
\providecommand{\BIBdecl}{\relax}
\BIBdecl

\bibitem{milek_eavesdropping_2018}
A.~Milek, E.~A. Butler, A.~M. Tackman, D.~M. Kaplan, C.~L. Raison, D.~A.
  Sbarra, S.~Vazire, and M.~R. Mehl, ``{Eavesdropping on Happiness Revisited: A
  Pooled, Multisample Replication of the Association Between Life Satisfaction
  and Observed Daily Conversation Quantity and Quality},'' \emph{Psychological
  Science}, vol.~29, no.~9, pp. 1451--1462, 2018.

\bibitem{wyatt2007conversation}
D.~Wyatt, T.~Choudhury, and J.~A. Bilmes, ``Conversation detection and speaker
  segmentation in privacy-sensitive situated speech data.'' in \emph{Proc.
  Interspeech}, 2007, pp. 586--589.

\bibitem{nelus2016towards}
A.~Nelus, S.~Gergen, J.~Taghia, and R.~Martin, ``Towards opaque audio features
  for privacy in acoustic sensor networks,'' in \emph{Proc. ITG Conf. on Speech
  Communication, VDE}, 2016, pp. 1--5.

\bibitem{seven}
P.~Thaine and G.~Penn, ``Extracting mel-frequency and bark-frequency cepstral
  coefficients from encrypted signals,'' in \emph{Proc. Interspeech}, 2019, pp.
  3715--3719.

\bibitem{federated}
M.~Hao, H.~Li, G.~Xu, S.~Liu, and H.~Yang, ``Towards efficient and
  privacy-preserving federated deep learning,'' in \emph{Proc. IEEE {Int.}
  {Conf.} on Communications}, 2019, pp. 1--6.

\bibitem{mehl2017ear}
M.~R. Mehl, ``{The Electronically Activated Recorder (EAR): A Method for the
  Naturalistic Observation of Daily Social Behavior},'' \emph{Current
  Directions in Psychological Science}, vol.~26, no.~2, pp. 184--190, 2017.

\bibitem{bitzer2016privacy}
J.~Bitzer, S.~Kissner, and I.~Holube, ``Privacy-aware acoustic assessments of
  everyday life,'' \emph{Journal of the Audio Engineering Society}, vol.~64,
  no.~6, pp. 395--404, 2016.

\bibitem{patino2021mcadams}
J.~Patino, N.~Tomashenko, M.~Todisco, A.~Nautsch, and N.~Evans, ``{Speaker
  Anonymisation Using the McAdams Coefficient},'' in \emph{Proc. Interspeech},
  2021, pp. 1099--1103.

\bibitem{raman2022conflab}
C.~Raman, J.~Vargas~Quiros, S.~Tan, A.~Islam, E.~Gedik, and H.~Hung, ``Conflab:
  A data collection concept, dataset, and benchmark for machine analysis of
  free-standing social interactions in the wild,'' \emph{Advances in Neural
  Information Processing Systems}, vol.~35, pp. 23\,701--23\,715, 2022.

\bibitem{voicepriv2020}
N.~Tomashenko, X.~Wang, E.~Vincent, J.~Patino, B.~M.~L. Srivastava, P.-G.
  No\'e, A.~Nautsch, N.~Evans, J.~Yamagishi, B.~O'Brien, A.~Chanclu, J.-F.
  Bonastre, M.~Todisco, and M.~Maouche, ``{The VoicePrivacy 2020 Challenge:
  Results and findings},'' \emph{Computer, Speech and Language}, vol.~74, 2022.

\bibitem{nakajima2018mosaic_speech}
Y.~Nakajima, M.~Matsuda, K.~Ueda, and G.~B. Remijn, ``{Temporal Resolution
  Needed for Auditory Communication: Measurement With Mosaic Speech},''
  \emph{Frontiers in Human Neuroscience}, vol.~12, 2018.

\bibitem{speechbrain}
M.~Ravanelli, T.~Parcollet, P.~Plantinga, A.~Rouhe, S.~Cornell, L.~Lugosch,
  C.~Subakan, N.~Dawalatabad, A.~Heba, J.~Zhong, J.-C. Chou, S.-L. Yeh, S.-W.
  Fu, and et~al., ``{SpeechBrain}: A general-purpose speech toolkit,'' 2021,
  arXiv:2106.04624.

\bibitem{CTC}
A.~Graves, S.~Fern{\'a}ndez, F.~Gomez, and J.~Schmidhuber, ``Connectionist
  temporal classification: labelling unsegmented sequence data with recurrent
  neural networks,'' in \emph{Proc. {Int.} {Conf.} on Machine Learning}, 2006,
  pp. 369--376.

\bibitem{ECAPA}
B.~Desplanques, J.~Thienpondt, and K.~Demuynck, ``{ECAPA}-{TDNN}: Emphasized
  channel attention, propagation and aggregation in {TDNN} based speaker
  verification,'' in \emph{Proc. Interspeech}, 2020, pp. 3830--3834.

\bibitem{librispeech}
V.~Panayotov, G.~Chen, D.~Povey, and S.~Khudanpur, ``Librispeech: An {ASR}
  corpus based on public domain audio books,'' in \emph{Proc. IEEE Int. Conf.
  on Acoustics, Speech and Signal Processing (ICASSP)}, 2015, pp. 5206--5210.

\bibitem{carletta_unleashing_2007}
J.~Carletta, ``Unleashing the killer corpus: experiences in creating the
  multi-everything {AMI} meeting corpus,'' \emph{Language Resources and
  Evaluation}, vol.~41, no.~2, pp. 181--190, 2007.

\bibitem{dawalatabad_ecapa-tdnn_2021}
N.~Dawalatabad, M.~Ravanelli, F.~Grondin, J.~Thienpondt, B.~Desplanques, and
  H.~Na, ``{ECAPA}-{TDNN} embeddings for speaker diarization,'' in \emph{Proc.
  Interspeech}, 2021, pp. 3560--3564.

\bibitem{VoxCeleb2}
J.~S. Chung, A.~Nagrani, and A.~Zisserman, ``{VoxCeleb2: Deep Speaker
  Recognition},'' in \emph{Proc. Interspeech}, 2018, pp. 1086--1090.

\bibitem{musan}
D.~Snyder, G.~Chen, and D.~Povey, ``{MUSAN}: A music, speech, and noise
  corpus,'' 2015, arXiv:1510.08484v1.

\bibitem{ko2017augment}
T.~Ko, V.~Peddinti, D.~Povey, M.~L. Seltzer, and S.~Khudanpur, ``A study on
  data augmentation of reverberant speech for robust speech recognition,'' in
  \emph{Proc. IEEE Int. Conf. on Acoustics, Speech and Signal Processing
  (ICASSP)}, 2017, pp. 5220--5224.

\bibitem{smeds_estimation_2015}
K.~Smeds, F.~Wolters, and M.~Rung, ``Estimation of {Signal}-to-{Noise} {Ratios}
  in {Realistic} {Sound} {Scenarios},'' \emph{Journal of the American Academy
  of Audiology}, vol.~26, no.~2, pp. 183--196, 2015.

\bibitem{jeub2009binaural}
M.~Jeub, M.~Schafer, and P.~Vary, ``A binaural room impulse response database
  for the evaluation of dereverberation algorithms,'' in \emph{Proc. IEEE Int.
  Conf. on Digital Signal Processing}, 2009, pp. 1--5.

\end{thebibliography}

\end{document}